\documentclass[prd,preprint,tightenlines,floatfix,showpacs,preprintnumbers,nofootinbib,eqsecnum]{revtex4}
\pdfoutput=1

 \usepackage[dvips,final]{graphicx}
  \usepackage{amssymb}
   \usepackage{amsmath}
    \usepackage{amsfonts}
     \usepackage{epsfig}
      \usepackage{bm} 
      \usepackage[dvipsnames,usenames]{color}
      \usepackage{comment}
      \usepackage{float}
      
      \usepackage[utf8]{inputenc}

\DeclareMathOperator\arctanh{arctanh}

\newcommand{\p}{_\perp}

\begin{document}

\title{Exclusive heavy quark-pair production in ultraperipheral collisions}

\author{Mateus Reinke Pelicer$^{1}$}
\email{m.reinke.pelicer@posgrad.ufsc.br}

\author{Emmanuel G.\ de Oliveira$^{1}$}
\email{emmanuel.de.oliveira@ufsc.br}

\author{Roman Pasechnik$^{1,2,3}$}
\email{Roman.Pasechnik@thep.lu.se}

\affiliation{
\\
{$^1$\sl Departamento de F\'isica, CFM, Universidade Federal 
de Santa Catarina, C.P. 476, CEP 88.040-900, Florian\'opolis, 
SC, Brazil
}\\
{$^2$\sl
Department of Astronomy and Theoretical Physics, Lund
University, SE-223 62 Lund, Sweden
}\\
{$^3$\sl Nuclear Physics Institute ASCR, 25068 \v{R}e\v{z}, 
Czech Republic\vspace{1.0cm}
}}

\begin{abstract}
\vspace{0.5cm}
In this article, we study the fully differential observables of exclusive production 
of heavy (charm and bottom) quark pairs in high-energy ultraperipheral $pA$ and 
$AA$ collisions. In these processes, the nucleus $A$ serves as an efficient source 
of the photon flux, while the QCD interaction of the produced heavy-quark pair with 
the target ($p$ or $A$) proceeds via an exchange of gluons in a color singlet state, described 
by the gluon Wigner distribution. The corresponding predictions for differential cross sections were obtained by using the dipole $S$-matrix in the McLerran-Venugopalan 
saturation model with impact parameter dependence for the nucleus target, and its recent generalization, for 
the proton target. Prospects of experimental constraints on the gluon Wigner distribution 
in this class of reactions are discussed.
\end{abstract}

\pacs{12.38.-t,12.38.Lg,12.39.St,13.60.-r,13.85.-t}

\maketitle

\section{Introduction}
\label{Sect:intro}

In QCD, the hadron structure is encoded in the so-called Wigner distributions \cite{Ji:2003ak,Belitsky:2003nz,Lorce:2011kd}.  These distributions are known to provide the most detailed information on 
the parton multi-dimensional imaging (tomography) in the target. The 5D Wigner distribution $W(x,\vec k_\perp,\vec b\p)$ depends on both the transverse momentum $\vec k\p$ of an exchanged parton and its impact parameter $\vec b\p$. While the Wigner distribution is in impact-parameter $\vec b\p$ representation, their Fourier transform as $\vec b\p \rightarrow \vec \Delta\p$ is known as the generalized transverse momentum distribution (GTMD) \cite{Meissner:2009ww,Hatta:2011ku,Lorce2013,Echevarria:2016mrc} in momentum representation.

These distributions are therefore sensitive to the angular correlation between $\vec b\p$ and $\vec k_\perp$ whose magnitude is determined by the elliptic Wigner distribution \cite{PhysRevLett.116.202301,Hagiwara:2016kam,Zhou:2016rnt}. It was shown earlier that the angular dependence of the Wigner distribution is particularly responsible for an elliptic flow in $pA$ collisions \cite{PhysRevD.95.094003,Hagiwara:2017ofm}, the angular correlations 
in deeply virtual compton scattering \cite{Hatta:2017cte} and in quasi-elastic photon-nucleus scattering \cite{Zhou:2016rnt}, etc. For a comprehensive review on the fundamental role of these distributions, see also Refs.~\cite{Boer:2011fh,Accardi:2012qut} and references therein. 

In Refs.~\cite{PhysRevLett.116.202301,ALTINOLUK2016373} considering an important example of electron-ion collisions in the high-energy limit 
it was demonstrated that the low-$x$ GTMD
\begin{equation}
xG( \vec k\p, \vec \Delta\p) \stackrel{x\rightarrow 0}{\approx}\, \frac{2 N_c}{\alpha_s} \left(k\p^2 - \frac{\Delta\p ^2}{4} \right) S(\vec k\p, \vec \Delta\p ) \,,
\end{equation}
is directly related to a Fourier transform of an impact parameter dependent forward dipole amplitude (or dipole $S$-matrix), $S(\vec r,\vec b)$ providing an important connection with the gluon saturation phenomena at low-$x$ (for a detailed review of the saturation effects and the Color Glass Condensate, see e.g. Ref.~\cite{Gelis:2010nm}).

Just like for lower-dimensional descendents, the collinear parton densities, the QCD perturbation theory cannot predict the key characteristics of the partial dipole amplitude $S(\vec k\p, \vec \Delta\p )$ and hence the gluon Wigner distributions, so potential possibilities for experimental measurements of such distributions or for setting constraints on them directly from the data gain large importance and have to be studied in detail \cite{PhysRevLett.116.202301}. The basic difficulty on the extraction of the Wigner distributions (or GTMDs) is typically associated with the fact that the differential cross section is not proportional to a GTMD itself but is given by its convolution integral with the light-cone wave function for a given projectile Fock state scattering off a target. Such integral is originated as a remnant of the loop integral in the exclusive production amplitude formed by the two exchanged gluons with the target (in the color-singlet state), and it is in general not analytically invertible.

A particular relevant class of scattering processes at hadron colliders, the high-energy ultraperipheral collisions (UPCs) with fully exclusive final states, provide essential means for 
accessing the hadron structure at relatively low momentum transfers and at low-$x$ due to both clean environment and complete reconstruction of kinematics of the exchanged gluons 
with a target. In UPCs, the relativistic systems scatter at typically large impact parameters 
by means of quasi-real Weisz\"acker-Williams (WW) photon exchange \cite{vonWeizsacker:1934nji,Williams:1934ad}. The WW flux is enhanced for heavy nucleus as the square of its charge making it an efficient source of photons. As was demonstrated recently in Ref.~\cite{PhysRevD.96.034009}, such gluon Wigner distribution can be constrained, or even directly extracted, from the data on exclusive 
light-quark di-jet photoproduction in the UPCs. Besides the largest contribution to the di-jet photoproduction signal, the use of light quarks in the final state as a source of di-jets is particularly 
beneficial as this channel enables to directly extract the gluon Wigner distribution the data on fully differential exclusive di-jet cross section. Indeed, the loop integral is {\it analytically} 
invertible in this particular case such that the components of the gluon Wigner distribution can be found as integrals of the components of the differential cross section.

However, a relevant non-trivial structure of the gluon Wigner distribution, in particular, its elliptic component, emerges when the transverse momenta of the produced $q$ and $\bar{q}$ 
are relatively low and do not significantly exceed the saturation scale of the process. This further prompts valid questions about the applicability of the QCD perturbation theory for reliable 
computation of the $\gamma+(gg)\to q\bar{q}$ matrix element in such a region of predominantly soft or semi-soft kinematics. As the only hard scale of this process is associated with the transverse
momentum of a produced jet (in dominant nearly back-to-back di-jet configurations), going to low (below a few GeV) jet transverse momenta is severely restricted by potentially large higher order 
effects, thus, limiting the capability of this method for a reliable extraction of the gluon Wigner distribution in the domain of its maximal enhancement and focussing on the tail high-$P_\perp$ 
regions only.

Another technical challenge is related to experimental capabilities for exclusive di-jet photoproduction measurements at high energies. Such a measurement requires reconstruction of full di-jet 
kinematics simultaneously triggering on large rapidity gap events only in order to suppress backgrounds causing the leakage of energy and transverse momentum into unreconstructed hadronic 
activity originating from the break up of the target, of the exchanged Pomeron and/or of the quasi-real photon. For this purpose, the forward proton (or nucleus) reconstruction is needed to ensure
exclusivity of the corresponding diffractive reaction. While ATLAS Forward Physics program offers certain possibilities for such a measurement, the acceptance in jet transverse momenta is highly
limited to high-$P_\perp$ kinematics only, with the lowest cut-off hardly going below 20 GeV or so.
\begin{figure}[H]
    \centering
    \includegraphics[width=.5\textwidth]{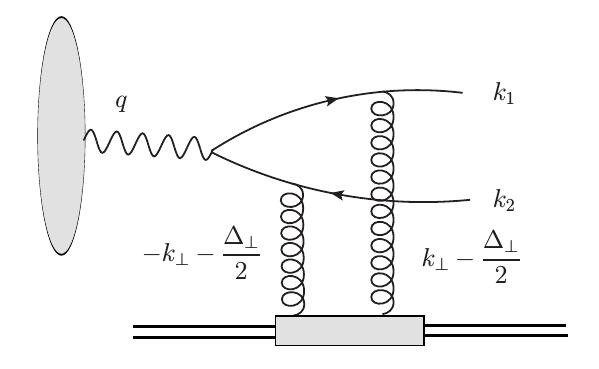} \hfill 
    \caption{Feynman diagram for a quark-pair photoproduction in UPCs. The projectile nucleus denoted by a shaded blob is a source of quasi-real WW photons. 
    The photon fluctuates into a $q\bar q$ dipole, which interacts by means of two-gluon exchange in a color singlet state with the target proton or nucleus.}
    \label{fig:Diagram}
\end{figure}

Such situation indicates that the use of heavy ($c$ and $b$) quarks, with measurements of exclusive open heavy flavor ($DD$ and $BB$ meson pairs) photoproduction, may help in resolving both
theoretical and experimental issues with probing the gluon Wigner distribution in the gluon saturation regime. First of all, it naturally provides a hard scale associated with the heavy quark mass,
thus, enabling the use of QCD perturbation theory, with less degree of uncertainty, even for a vanishing quark transverse momentum. Second, reconstruction of heavy-flavored mesons can be 
performed at much lower transverse moments than for jets and, despite of a smaller cross section, with a much better control of QCD backgrounds. As a price to pay for such an improvement,
for heavy quarks the convolution integral in the diffractive amplitude is no longer analytically invertible, so for now we can only make predictions for the corresponding observables in the framework
of a given model for the gluon Wigner distribution.

These arguments motivate us to perform the first detailed study of exclusive $c\bar{c}$ and $b\bar{b}$ pairs photoproduction in UPCs in fully resolved kinematics when the target survives the interaction and is detected 
in a forward region as shown in Fig.~\ref{fig:Diagram}. Such process is driven by at least the exchange of 
two gluons in a color singlet state, to the leading order in QCD perturbation theory. This study is performed for 
a large nucleus target in the framework of the recently upgraded McLerran-Venugopalan (MV) incorporating the gluon saturation 
phenomenon with impact parameter dependence \cite{PhysRevD.49.3352, PhysRevD.95.094003}. For the case of proton target, we have employed a recent generalization of 
the MV model presented in Ref.~\cite{PhysRevD.95.094003}. We analyze the corresponding observables particularly sensitive to the elliptic gluon Wigner distribution in the physically relevant regions of exchanged gluon 
transverse momenta close to the saturation scale. While a direct reconstruction of the Wigner distribution from 
the data remains a challenging task, we identify certain non-trivial behavior in the differential cross section directly related to the features of the elliptic distribution.

The paper is organized as follows. In Section~\ref{Sect:Formalism}, the light cone approach for heavy quark exclusive photoproduction is formulated in terms of the dipole $S$-matrix, and that is a new analytic results as far as we know. In Section~\ref{sec:MV-model}, we provide a short description of the MV model for both the large nucleus target and the proton target. Section~\ref{Sect:Results} contains numerical results for the structure functions for the massive and massless cases as well as the production cross sections for $c\bar{c}$ and $b\bar{b}$ in the case of lead and protons targets. Finally, concluding remarks and a summary are given in Section~\ref{Sect:Concl}.

\section{Light-cone dipole approach for exclusive heavy-quark pair photoproduction in UPCs}
\label{Sect:Formalism}

\subsection{Kinematics}

We work in the nucleus-target center of mass frame. The photon is collinear to the $z$-axis direction, and carries energy $\omega$. Since we are working with quasi-real photons ($q^2 \approx 0$), its momentum can be written as $q = (\sqrt{2} \omega, 0, 0\p)$ in Sudakov (light-cone) variables and the longitudinal polarization contribution to the cross section is negligible. 

The incoming gluon transverse momenta are denoted as $\vec k_\perp - \vec \Delta_\perp/2$ and $- \vec k_\perp - \vec \Delta_\perp/2$, where $k\p$ is integrated as the loop momentum.  Their energy and longitudinal momenta are neglected in the limit of $x \rightarrow 0$. In what follows, we do not consider QCD evolution in $x$ variable (or rapidities). Instead, we perform our analysis in the forward kinematics, such that the gluonic contribution to the quark longitudinal momentum is small, and we justify neglecting it. In this approximation the target contributes only with a total transverse momentum $-\vec \Delta_\perp$ to the final $q \bar q$ pair. 

The final-state heavy quarks studied here are charm and bottom quarks, with masses $m_c = 1.4$\,GeV and $m_b = 4.7$\,GeV, respectively. The quark will carry momentum fraction $z$ of the projectile photon and has transverse momentum $ -\vec P\p + \vec k\p$, coming from the photon splitting, while the antiquark will carry momentum fraction $(1-z)$ from the photon and have transverse momentum $\vec P\p - \vec k\p$. After the di-gluon exchange, the quark will acquire the following light-cone momentum components
 \begin{align*}
k_1^+ = z \sqrt{2} \omega, \; \; \vec k_{1 \perp} = -\vec P\p - \frac{\vec \Delta\p}{2},  \end{align*}
and, analogously, for the antiquark
\begin{align*}
 k_2^+ = (1-z) \sqrt{2} \omega, \;\; \vec k_{2 \perp} = \vec P\p - \frac{\vec \Delta\p}{2}. 
 \end{align*}
The $k_i^-$ (where $i=1,2$) components are determined from the condition that the final states are on mass shell. If the quark rapidities $y_i=$ln$(\sqrt{2}k_i^+/\sqrt{k_{i \perp}^2 + m_Q^2})$ and transverse momenta are measured, then the quark momentum fraction $z$ and the photon energy $\omega$ are fixed. 

\subsection{Exclusive $q\bar q$ photoproduction cross section}

The cross section for a UPC between a projectile nucleus $A$ and a target, which can be either a nucleus or a proton, i.e. $j=A,p$, can be written as 
\begin{equation}
\frac{d\sigma ^{A j} }{d y_1 \, d y_2 \, d^2 \vec P\p \, d^2 \vec \Delta\p} =
\omega \frac{d N_\gamma}{d\omega} \frac{d\sigma ^{\gamma j} }{d y_1 \, d y_2 \, d^2 \vec P\p \, d^2 \vec \Delta\p} \, ,
\end{equation}
where $\omega dN_\gamma/d\omega$ is the photon number density. These quasi-real photons coming from the projectile heavy ion are modeled by the WW photon distribution \cite{vonWeizsacker:1934nji,Williams:1934ad,jackson_classical_1999}
\begin{equation}
 \frac{d N_\gamma}{d\omega} = \frac{2 Z^2 \alpha}{\pi \omega} \left[ \xi_{jA} K_0(\xi_{ jA}) K_1(\xi_{jA}) -\frac{\xi_{ja}^2}{2} \left(K_1^2(\xi_{jA}) - K_0^2 (\xi_{jA} ) \right) \right].
 \end{equation}
In the above expression, $Z$ is the atomic number of the projectile, $\alpha$ is the fine structure constant, and $\xi_{jA} = \omega (R_j + R_A)/\gamma$ is defined in terms of the Lorentz factor $\gamma = \sqrt{s_{jA}}/2M_p$, the target and nucleus radii, $R_j$ and $R_A$, respectively, and
the $jA$ center-of-mass energy, $\sqrt{s_{jA}}$. The $R_j + R_A$ dependence guarantees that the photons can only interact with the target when there is no overlap between the projectile and the target in impact parameter space. For a review on peripheral collisions and photon fluxes, see Ref.~\cite{BALTZ20081}.  

The partonic cross section was calculated using the light-cone Feynman rules \cite{PhysRevD.50.3134, Kovchegov:2012mbw}. The photon-splitting into a $q\bar q$ dipole can be calculated analytically via the photon wave-function, while the di-gluon-dipole interaction is encoded in the transition matrix $T$. The latter is defined as $T = 1 - S$ in terms of the dipole $S$-matrix describing the quark-antiquark dipole scattering off the target and will be discussed in the following sections. For the parton-level cross section, we have the following expression
\begin{equation}
\label{cross-section}
\frac{d\sigma ^{\gamma j} }{d\mathcal{PS}} = 2(2 \pi ) ^2 N_c \alpha e_q ^2 z (1-z)   \left[ (z^2 + (1-z)^2) | \mathcal{\vec M}_0 | ^2   + (m_Q^2 /P\p^2) |\mathcal{\vec M}_1 | ^2 \right],
\end{equation}
where the phase space element is given by $d\mathcal{PS} = d y_1 dy_2 d^2\vec P\p d^2 \vec \Delta\p$. The functions $\mathcal{\vec M}_0$ and $\mathcal{\vec M}_1$ are expressed in terms of the $T$-matrix as follows
\begin{equation}\label{matrixel0}
\mathcal{\vec M}_0 ( \vec P\p , \vec \Delta \p) = \int \frac{d^2\vec k\p}{2\pi} \frac{\vec P\p - \vec k\p}{(\vec P\p - \vec k \p)^2 + m_Q^2}  T(\vec k\p, \vec \Delta \p ),
\end{equation}
and
\begin{equation}\label{matrixelm}
\mathcal{\vec M}_1 ( \vec P\p , \vec \Delta \p) = \int \frac{d^2\vec k\p}{2\pi} \frac{\vec P\p}{(\vec P\p - \vec k \p)^2 + m_Q^2} T(\vec k\p, \vec \Delta \p ) .
\end{equation}
Formally, in the above expression, it is $S$ that should be in place of $T$. However, we took out the non-interaction term 1 from the dipole $S$-matrix, so for vanishing $P\p$ we will have $\mathcal{M}_0 = \mathcal{M}_1= 0$ (see e.g. Ref.~\cite{PhysRevD.96.034009}).

\subsection{$T$ matrix}

It has been shown that azimuthal angular correlations between $\vec k\p$ and $\vec \Delta \p$ are important in the $T$-matrix when working in the saturation regime \cite{bartels_azimuthal_1996, Diehl:1996st, ALTINOLUK2016373}. In the limit where $P\p \gg \Delta\p$, these correlations can be taken into account by performing an expansion in Fourier harmonics 
\begin{equation}
T(\vec k \p, \vec \Delta \p)  = T_0(k\p, \Delta\p) + \cos 2(\phi_k - \phi_\Delta) T_\epsilon(k\p, \Delta\p) + \cdots
\end{equation}
The main contribution to the cross section comes from the isotropic part $T_0$, and we have a sub-leading contribution from the elliptic part $T_\epsilon$ \cite{PhysRevLett.116.202301}. The latter is the subject of our further analysis, while the higher order harmonics are suppressed.

Since the $T$ matrix contains information on the strong interaction in the saturation regime, it is unfeasible to use pQCD to calculate it, and we must utilize a physically reasonable model for it that should incorporate the gluon saturation effects. Typically, such models are formulated in impact parameter space, where we define $\vec r\p$ to be the dipole size and $\vec b\p$ -- the impact parameter in the transverse plane. To connect the model with our representation in momentum space, we define the Fourier transform as
\begin{equation}\label{Tfourier}
T(\vec k\p, \vec \Delta\p) =   \int \frac{d^2 \vec b\p}{(2 \pi)^2}   \int \frac{d^2 \vec r\p}{(2 \pi)^2}  e^{- i \vec k\p \cdot r\p} e^{-i \vec \Delta\p \cdot \vec b\p }e^{- \epsilon_b b\p^2} e^{- \epsilon_r r\p^2} T(\vec r\p, \vec b\p).
\end{equation}
This integral can be related to the isotropic and elliptic contributions by expanding the Fourier exponentials in Bessel functions of first kind 
according to the following relation 
\begin{equation}
e^{i x \cos \phi} =  \sum\limits_{n=-\infty}^\infty i^n J_n(x) e^{i n \phi} \,.
\end{equation}
The isotropic part is related to $n=0$, while the elliptic term appears for $n=\pm 2$. The odd terms do not contribute to UPCs, but it should be noted that they can be of importance in other processes where the nucleus exhibits an inhomogeneity in the radial direction  \cite{Lorce2013, Boer2018}. As a result, we obtain
\begin{equation}\label{T0_conv}
T_0(k\p, \Delta\p) =   \int \frac{d^2 \vec b\p}{(2 \pi)^2}   \int \frac{d^2 \vec r\p}{(2 \pi)^2}  J_0(k\p r\p) J_0(\Delta\p b\p ) e^{- \epsilon_b b\p^2} e^{- \epsilon_r r\p^2} T(\vec r\p, \vec b\p),
\end{equation}
\begin{equation}\label{Te_conv}
T_\epsilon(k\p, \Delta\p)=  2 \int \frac{d^2 \vec b\p}{(2 \pi)^2}   \int \frac{d^2 \vec r\p}{(2 \pi)^2}   J_2(k\p r\p) J_2(\Delta\p b\p)e^{- \epsilon_b b\p^2} e^{- \epsilon_r r\p^2} \cos 2 (\phi_b - \phi_r) T(\vec r\p, \vec b\p).
\end{equation}
Here, the Gaussian-type exponentials act as dumping terms to improve the convergence of the integrals, which are highly oscillatory on the periphery. Physically, they provide a physical cut-off accounting for confinement effects, therefore, the $\epsilon$ parameters are inversely related to the typical size of the bound systems (nucleon and/or nucleus) and thus should be small compared to the hard scale of the process. We expect the $b\p$ parameter to be smaller than the target size, such that $\epsilon_b = 1/R_j^2$. We also set $\epsilon_r = (0.5\,$fm$)^{-2}$ as the photon splitting into the quark-antiquark pair will be suppressed when the transverse separation within the pair is larger than a typical hadron length-scale.

\subsection{The structure functions in the massive quark case}

Since the angular dependence is explicit, we can calculate analytically the azimuthal integrals in Eqs.~\eqref{matrixel0} and \eqref{matrixelm}. The first expression is then reduced to:
\begin{equation}
\mathcal{\vec M}_0 = \frac{\vec P\p}{P\p^2} \left[ A(P\p, \Delta\p) + B (P\p, \Delta\p) \cos 2(\phi_P - \phi_\Delta) \right] \,,
\end{equation}
which is now written in terms of the following two structure functions
\begin{align}
A(P\p, \Delta\p) =& \int_0^\infty k\p dk\p  \frac{P\p^2}{k\p^2 + P\p^2 + m_Q^2 + \sqrt{(k\p^2 + P\p^2 + m_Q ^2)^2 - 4 P\p^2 k\p^2} }  \\
& \times\left[ 1 + \frac{P\p^2 + m_Q^2 - k\p^2 }{\sqrt{(k\p^2 + P\p^2 + m_Q ^2)^2 - 4 P\p^2 k\p^2} } \right]T_0(k\p, \Delta\p), \nonumber
\end{align}
and
\begin{align}
B(P\p, \Delta\p) =& \frac{1}{2 P\p^2} \int_0^\infty \frac{dk\p}{k\p}   (P^2\p - k^2\p - m_Q^2) T_\epsilon (k\p, \Delta\p) \nonumber \\
&\times  \left[  \frac{ (k\p^2 + P\p^2 + m_Q ^2)^2 - 2 k\p^2 P\p^2}{\sqrt{(k\p^2 + P\p^2 + m_Q ^2)^2 - 4 P\p^2 k\p^2}}   - (P\p^2 + k\p^2 + m_Q^2) \right] \,.
\end{align}
These generalize the results of Ref.~\cite{PhysRevD.96.034009} to the massive case. On top of that one should consider two additional structure functions $C$ and $D$ entering the amplitude in Eq.~\eqref{matrixelm}, in proportion to the quark mass, such that
\begin{equation}
\mathcal{\vec M}_1 = \frac{\vec P\p}{P\p^2} \left[ C(P\p, \Delta\p) + D (P\p, \Delta\p) \cos 2(\phi_P - \phi_\Delta) \right] ,
\end{equation}
where
\begin{equation}
C(P\p, \Delta\p) =\int_0^\infty k\p dk\p \frac{P\p^2}{\sqrt{(k\p^2 + P\p^2 + m_Q ^2)^2 - 4 P\p^2 k\p^2}}  T_0(k\p, \Delta\p),
\end{equation}
\begin{align}
D(P\p, \Delta\p) & = \int_0^\infty \frac{dk\p}{k\p} \left[k\p^2 + P\p^2 + m_Q^2- \frac{ (k\p^2 + P\p^2 + m_Q ^2)^2 - 2 P\p^2 k\p^2}{\sqrt{(k\p^2 + P\p^2 + m_Q ^2)^2 - 4 P\p^2 k\p^2}}  \right] T_\epsilon (k\p, \Delta\p).
\end{align}
The $C$ and $D$ are defined so that all structure functions have the same mass dimension $[A]=[B]=[C]=[D]=[{\rm mass}]^{-2}$. Starting from the parton-level cross section in Eq.~\eqref{cross-section}, the above results enable us to represent the hadron-level cross section as follows
\begin{align}
\frac{d\sigma^{Aj} }{d\mathcal{PS}} =\, & \omega \frac{d N}{d\omega} 2(2 \pi ) ^2 N_c \alpha_{em} e_q ^2 z (1-z) \frac{1}{P\p^2}\nonumber \\
&\left\{ (z^2 + (1-z)^2) \; \left[ A(P\p, \Delta\p) + B (P\p, \Delta\p) \cos 2(\phi_P - \phi_\Delta) \right]^2 \right.  \nonumber \\ 
& \left. + \frac{m_f^2}{P\p^2} \; \left[ C(P\p, \Delta\p) + D (P\p, \Delta\p) \cos 2(\phi_P - \phi_\Delta) \right]^2 \right\} \, .
\end{align}

\section{Phenomenological McLerran--Venugopalan model for the dipole $S$ matrix}
\label{sec:MV-model}

In the MV model \cite{PhysRevD.49.3352, PhysRevD.50.2225}, a heavy nucleus can be treated as a semi-classical color field, where the gluons have a high occupation number which is controlled by the saturation scale $Q_{s}$. In a recent proposal \cite{PhysRevD.95.094003}, Iancu and Rezaeian generalize
the MV formalism for the target in order to incorporate a non-trivial impact-parameter dependence following the other saturation models such as IP-Sat and bCGG \cite{Kowalski:2003hm, PhysRevD.78.014016}. Such a generalization enables to analyze azimuthal asymmetries in heavy-ion collisions as a consequence of the collective phenomena in the initial state. In what follows, we employ this model in studies of heavy-quark photoproduction in UPCs. We will work in the picture where the dipole experiences multiple soft scatterings off the target. The $T$-matrix is, 
in the Glauber approximation,
\begin{equation}
T(\vec b\p, \vec r\p) = 1 - \exp\left(- A \, N (\vec b\p, \vec r\p) \right),
\end{equation}
where $N(\vec b\p, \vec r\p)$ is the single dipole scattering amplitude. 
We use $A=1$ for the proton target. The angular correlations in configuration space are generated by expanding the single scattering amplitude in Fourier harmonics, as done in the previous section 
$N(\vec b\p, \vec r\p) = N_0( b\p, r\p) +\cos 2\delta \phi_{rb} \, N_\epsilon( b\p, r\p)$. 

For a nucleus target, one gets \cite{PhysRevD.95.094003}
\begin{align}
N_0(b\p, r\p)  = \pi R^2 Q_{0, s}^2 r\p^2\text{ln} \left(\frac{1}{r\p^2 m_g^2} + e \right) & \left[T_A(b\p) + R^2 \left(T_A'' (b\p) + \frac{1}{b\p} T_A'(b\p) \right) \right] \nonumber \\
&+ \frac{\pi R^2}{3 m_g^2} Q_{0, s}^2 r\p^2 \left( T_A''(b\p) + \frac{1}{b\p} T_A'(b\p) \right),
\end{align}
and
\begin{equation}
N_\epsilon(b\p, r\p) = \frac{\pi R^2}{6 m_g^2} Q_{0, s}^2 r\p^2 \left(T_A'' (b\p) - \frac{1}{b\p} T_A'(b\p) \right).
\end{equation}
In the above equations, the parameter $m_g$ is an IR regulator, which acts as an effective gluon mass. We set it to $0.25$ GeV. We define the saturation scale at zero impact parameter as $Q_{0,s} = 1/R$, where $R$ is a scale related to the width of the proton color-charge distribution. We use $R=2$ GeV$^{-1}$, based on best fit values obtained in different saturation models to the HERA and NMC data \cite{PhysRevD.87.034002, PhysRevD.88.074016}. The proton radius is fixed as $R_p = 0.8$ fm, whilst that of the nucleus is given by $R_A = (1.12$ fm$)A^{1/3}$. The $T_A$ is the thickness function of the nucleus target found as follows 
\begin{equation}
T_A(b\p) = \int dz \rho_A\left(\sqrt{b\p^2 + z^2}\right),
\end{equation}
where $\rho_A(r)$ is the Woods-Saxon nuclear density distribution, $\rho_A(\vec r) = N_A \,\left[1+ \exp\left((r - R_A)/\delta \right) \right]^{-1}$, and $N_A$ is determined by the normalization condition $ \int d^3 \vec r \rho_A(\vec r) = 1$. In numerical analysis, we use $\delta=0.54$\,fm.
 \begin{figure}[htb]
    \centering
    \includegraphics[width=.48\textwidth]{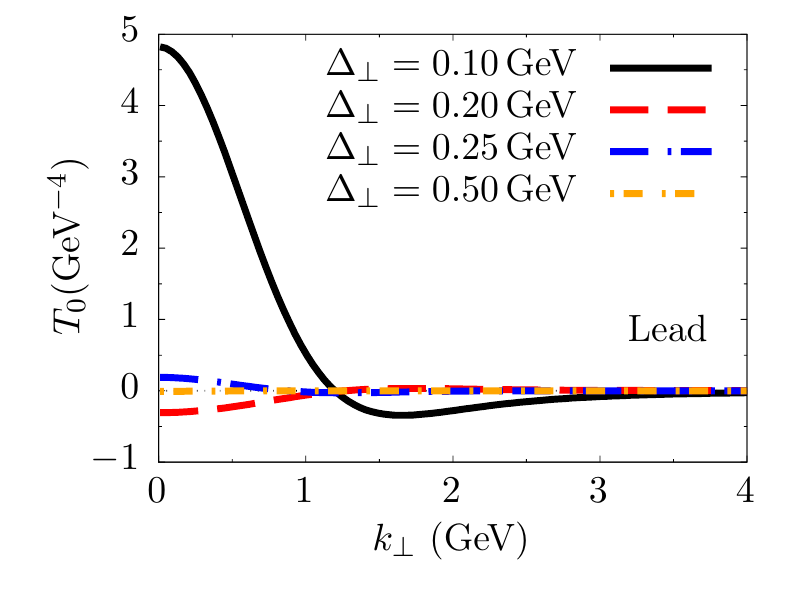} \hfill 
    \includegraphics[width=.48\textwidth]{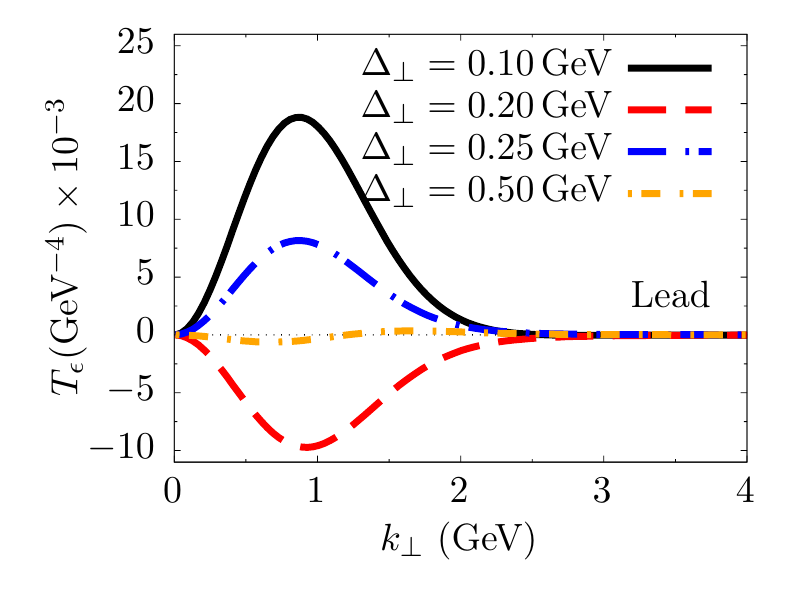}
    \caption{Left: Isotropic component of the $T$-matrix for lead (Pb) nucleus. We fix the values of $\Delta\p$ at $0.1$, $0.2$, $0.25$ and $0.5$ GeV, and make the plot as a function of the gluons transverse momentum $k\p$. Right: The same plot is made for the elliptic component.}
    \label{fig:T-nucleus}
\end{figure}

The $T$-matrix for a lead (Pb, $A=208$) nucleus is plotted in Fig.~\ref{fig:T-nucleus}. The isotropic part falls very fast with $\Delta\p$ and has a larger contribution from soft gluon momentum ($k\p \lesssim 2$ GeV). The elliptic part falls slower, becoming more important at higher $\Delta\p$. Note that its peak is kept at an almost constant $k\p$, as expected, since this is controlled by the saturation scale.

For a proton target, the single scattering amplitudes are given by an integral over the relative transverse momentum between the soft gluons $p\p$, and we have evaluated analytically one of the integrals given in Ref.~\cite{PhysRevD.95.094003}. Namely,
\begin{align}\label{eq:N0p}
N_0(b\p, r\p) =& \frac{Q_{0, s}^2 r\p^2}{4}  e^{-b\p^2/4R^2} \text{ln} \left( \frac{1}{r\p^2 m_g^2} + e\right) + Q_{0, s}^2 R^2 r\p^2 \int_0^\infty p\p d p\p e^{- p\p^2 R^2} \nonumber \\
& \times J_0(b\p p\p) \frac{p\p \sqrt{p\p^2 + 4 m_g^2} - 2 (p\p^2 + 2m_g^2) \arctanh{\left( \frac{p\p}{\sqrt{p\p^2 + 4 m_g^2}} \right)} }{2 p\p \sqrt{p\p^2 + 4 m_g^2}},
\end{align}
\begin{align}
N_\epsilon(b\p, r\p) =  Q_{0, s}^2 R^2 r\p^2 \int_0^\infty p\p d p\p & e^{- p\p^2 R^2}  J_2(b\p p\p) \nonumber \\
&\times \frac{p\p \sqrt{p\p^2 + 4 m_g^2} - 4 m_g^2 \arctanh{\left( \frac{p\p}{\sqrt{p\p^2 + 4 m_g^2}} \right)} }{2 p\p \sqrt{p\p^2 + 4 m_g^2}}.
\end{align}
The Gaussian-like exponentials in impact parameter, $b\p$, and relative momentum, $p\p$, are introduced in the model due to the color distribution in the transverse plane of the proton. The first term of Eq.~\eqref{eq:N0p} is the usual one presented in the original MV model~\cite{PhysRevD.49.3352}.
\begin{figure}[htb]
    \centering
    \includegraphics[width=.48\textwidth]{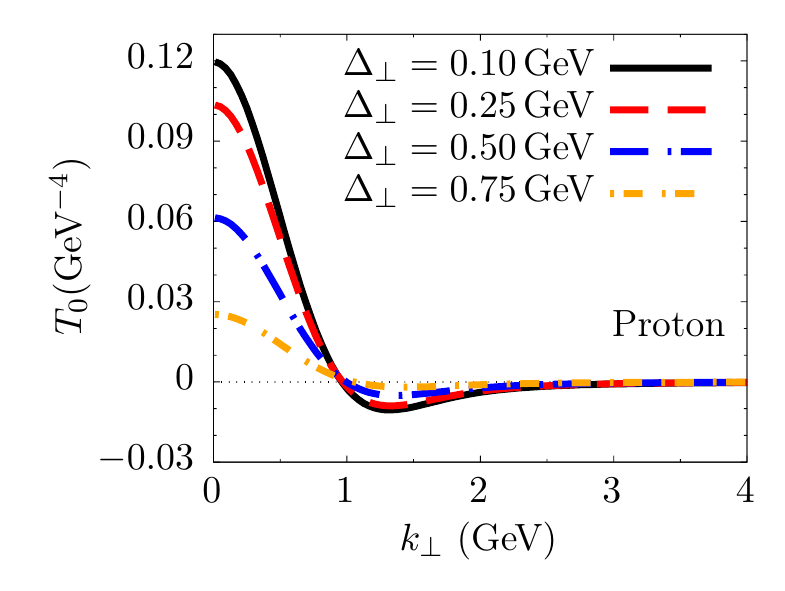} \hfill 
    \includegraphics[width=.48\textwidth]{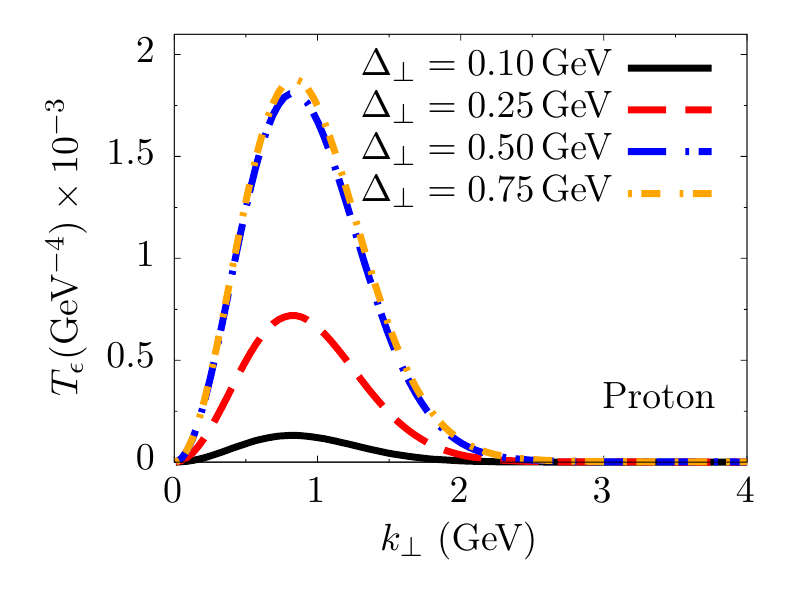}
    \caption{Same as in Fig.~\ref{fig:T-nucleus}, but for a proton target instead of a nucleus.}
    \label{fig:T-proton}
\end{figure}

We present the $T$-matrix for the proton in Fig.~\ref{fig:T-proton}. While the isotropic component falls much slower then in the case of a nucleus target, the elliptic one rises up to $\Delta\p \approx 0.7$\,GeV, and then starts to decrease. 

\begin{figure}[!h]
    \centering
    \includegraphics[width=.48\textwidth]{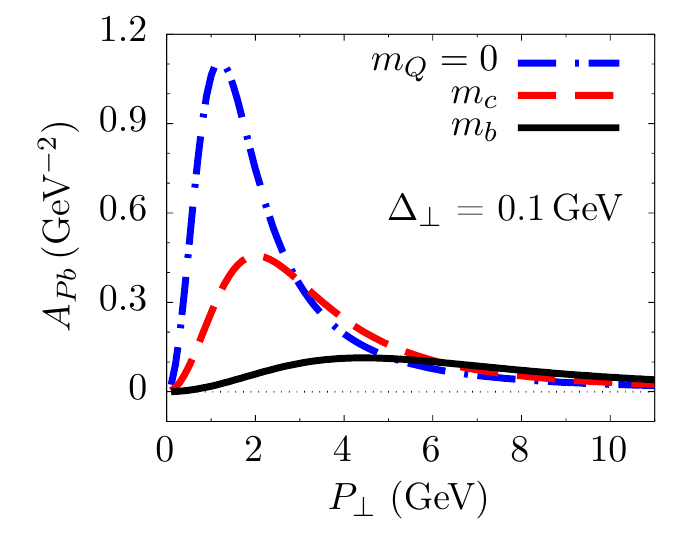} \hfill 
    \includegraphics[width=.48\textwidth]{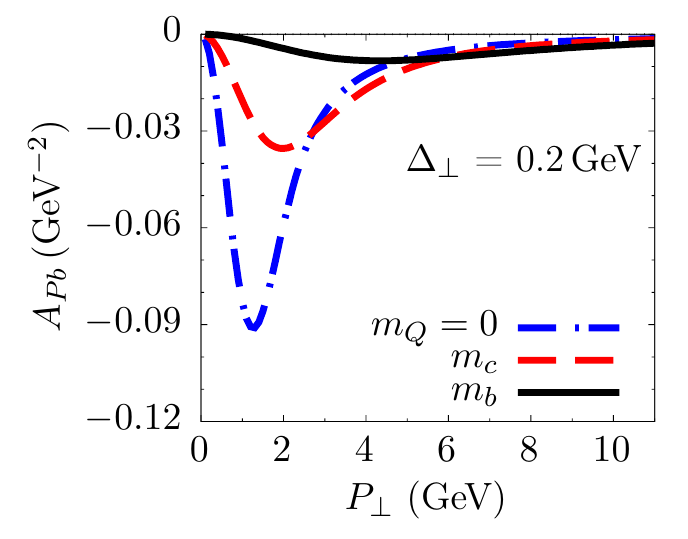}
    \caption{The structure function $A$ for the lead (Pb) nucleus target for heavy ($c,b$) and massless quarks in as a function of $P\p$. Two values of $\Delta\p=0.1$ and $0.2$ GeV are shown on left and right panels, respectively. As the quark mass increases, the peak of 
    the structure function gets smaller and moves to larger $P\p$. }
    \label{fig:A}
\end{figure}

\section{Numerical results}
\label{Sect:Results}

\subsection{Structure functions}

\begin{figure}[!h]
    \centering
    \includegraphics[width=.48\textwidth]{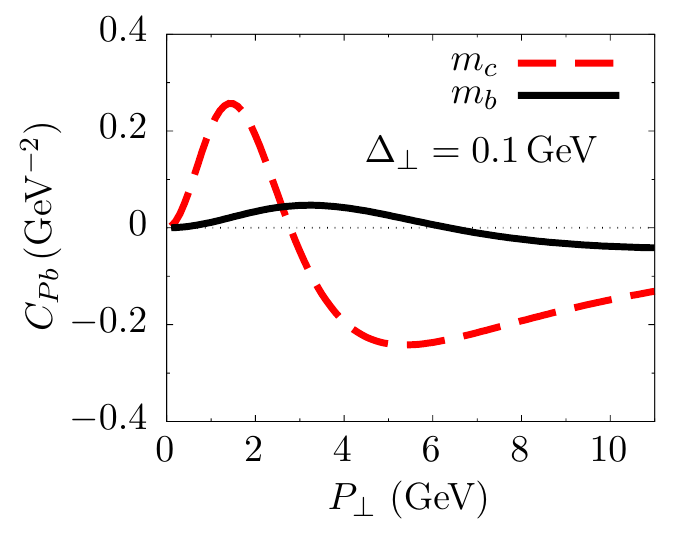} \hfill 
    \includegraphics[width=.48\textwidth]{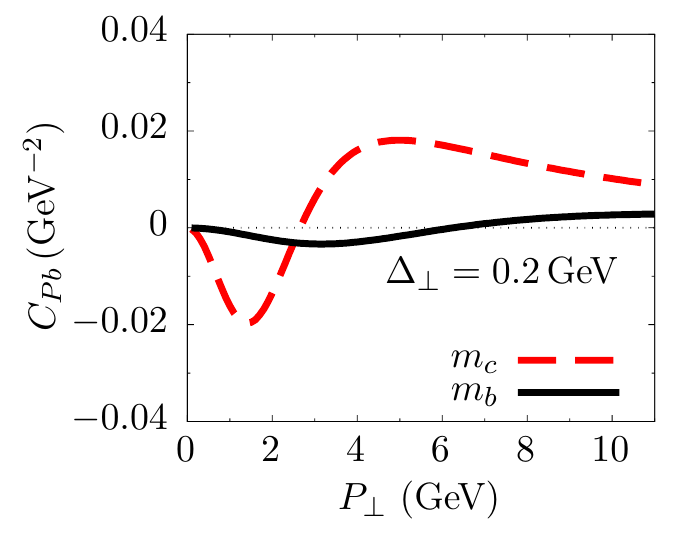}
    \caption{Same as Fig.~\ref{fig:A}, but for the $C$ structure function.}
    \label{fig:C}
\end{figure}

\begin{figure}[htb]
    \centering
    \includegraphics[width=.48\textwidth]{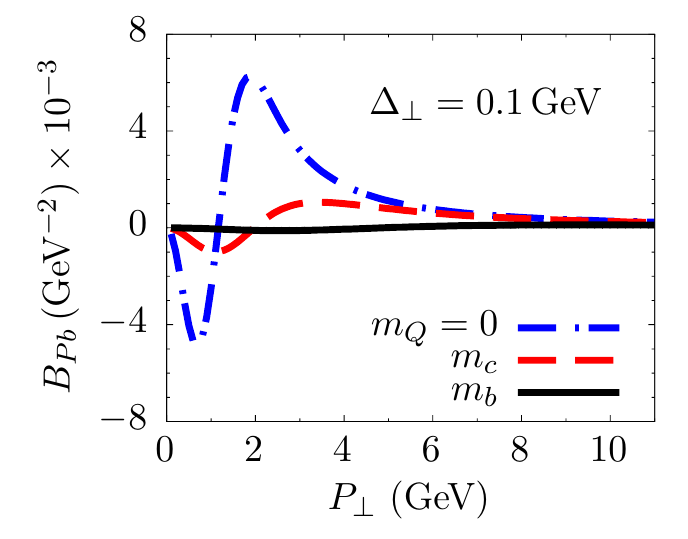} \hfill 
    \includegraphics[width=.48\textwidth]{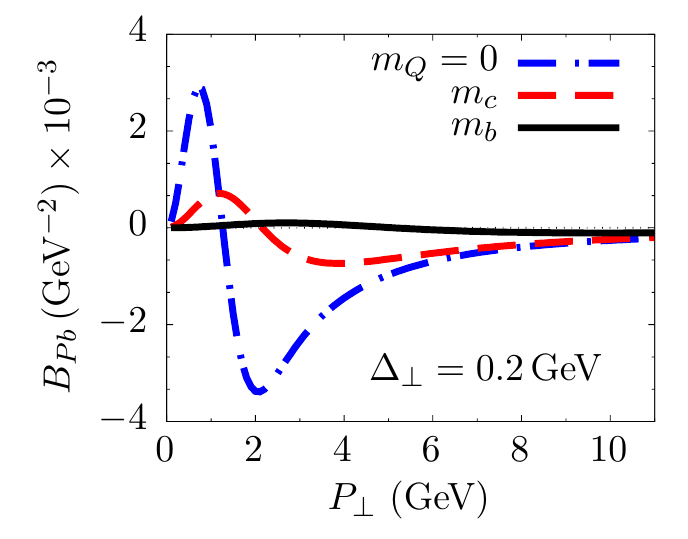}
    \caption{Same as Fig.~\ref{fig:A}, but for the $B$ structure function.}
    \label{fig:B}
\end{figure}

\begin{figure}[htb]
    \centering
    \includegraphics[width=.48\textwidth]{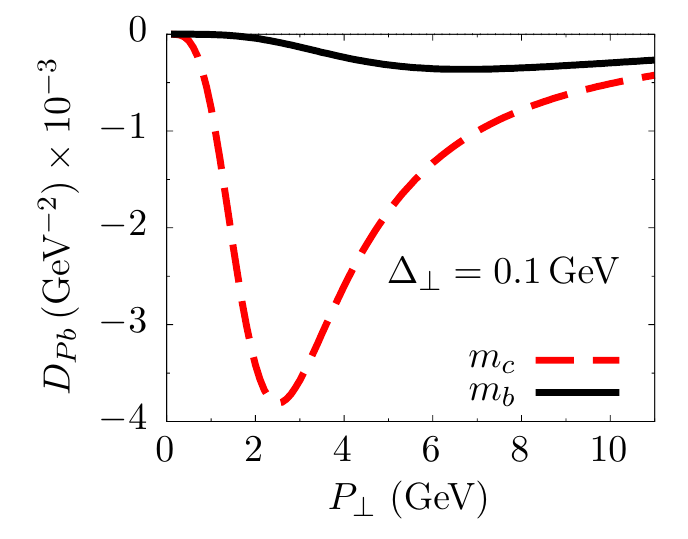} \hfill 
    \includegraphics[width=.48\textwidth]{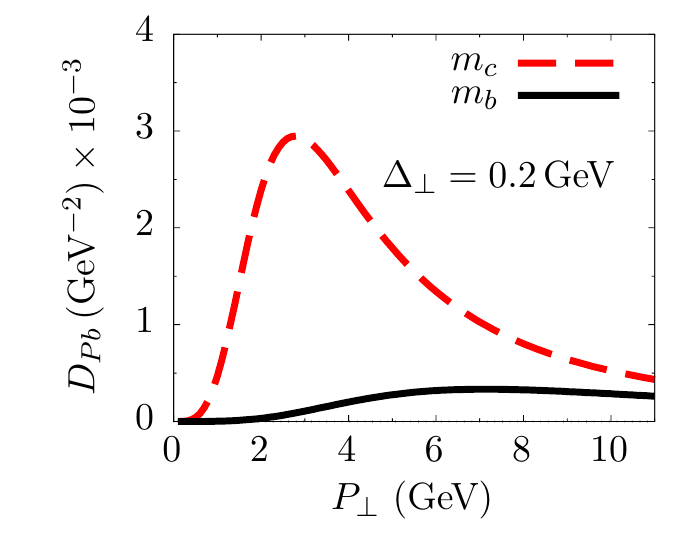}
    \caption{Same as Fig.~\ref{fig:A}, but for the $D$ structure function.}
    \label{fig:D}
\end{figure}

As one of the main new results of our analysis, we show the structure functions $A,C,B,D$ in Figs.~\ref{fig:A}, \ref{fig:C}, \ref{fig:B}, \ref{fig:D}, respectively, for heavy ($c,b$) quarks exclusively produced off a lead (Pb) nucleus target as functions of the quark-antiquark relative transverse momentum, $P\p$. Since the $A,B$ structure functions are present also for massless quark case, previously studied in Ref.~\cite{PhysRevD.96.034009}, these are shown by dashed-dotted curves in addition to the corresponding ones for heavy quarks. These results are shown for two distinct values of $\Delta\p=0.1$ and $0.2$\,GeV,  on left and right panels, respectively. Due to the presence of the additional structure functions, the information that can be obtained by probing the nucleus structure by means of heavy quarks is clearly richer than for the case of massless quarks.

When analyzing absolute values, one notices that the structure function $C$ is comparable to $A$, both determined in terms of the isotropic Wigner distribution, as is $D$ to $B$, given in terms the elliptic Wigner distribution. Therefore, the introduction of quark masses is not a matter of only correcting the standard $A$ and $B$ functions but surely the new $C$ and $D$ structure function considered here for the first time must be included. 

Figs.~\ref{fig:A}--\ref{fig:D} show that, as quark mass increases, absolute values of the peaks of the structure functions get smaller. This behavior is similar to a rescaling of $P\p$ accompanying a ``stretch'' of the shape of the distributions along the horizontal axis such that the decreasing (due to a mass suppression) peaks of the structure functions move to larger $P\p$. More specifically, in the large quark mass limit, the position of the peaks approximately scales with the quark mass as $P\p^{\rm peak}\sim m_Q$, while this dependence is violated for small quark masses.

The effect of going from $\Delta\p=0.1$ to $0.2$ GeV is basically a change of the sign and a reduction of the absolute value. However, for larger $\Delta\p$, the elliptic structure functions $B$ and $D$ become more relevant when compared to $A$ and $C$, as the latter have a stronger reduction in absolute values. These features are a direct consequence of oscillations and the general behavior of the $T_0$ and $T_\epsilon$ functions in the MV model, as it is shown in Fig.~\ref{fig:T-nucleus}. It is important to notice that the $P\p$ values of the peaks are kept almost constant for the different $\Delta\p$. This is the same as for the $T$-matrix: the peaks are controlled by the saturation scale.
\begin{figure}[htb]
    \centering
    \includegraphics[width=.48\textwidth]{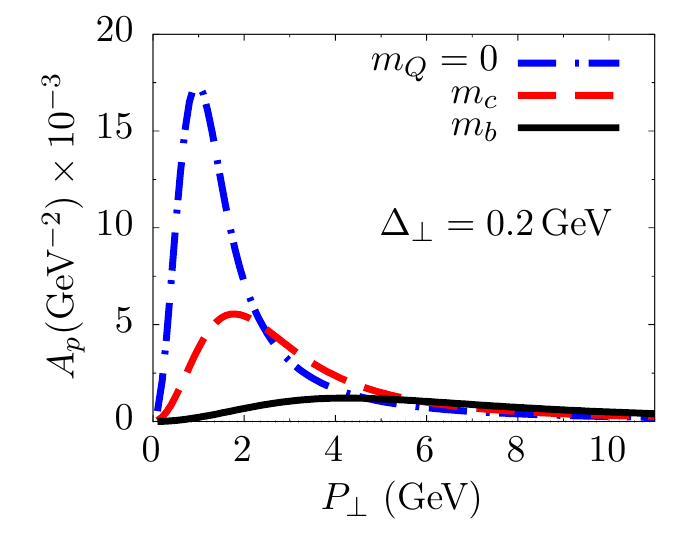} \hfill 
    \includegraphics[width=.48\textwidth]{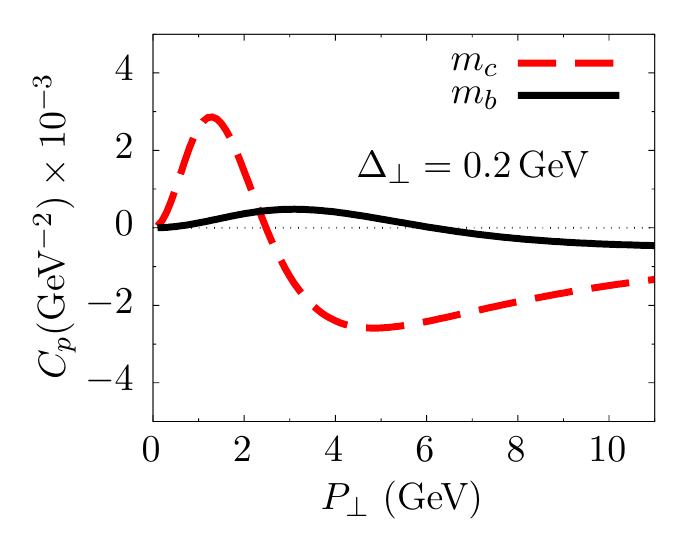}
    \caption{Left: the structure function $A$. Right: the structure function $C$. Both cases correspond to the proton target and are shown for massless and heavy ($c,b$) quarks as functions of $P\p$ for a fixed $\Delta\p=0.2$ GeV.}
    \label{fig:proton:AC}
\end{figure}

\begin{figure}[htb]
    \centering
    \includegraphics[width=.48\textwidth]{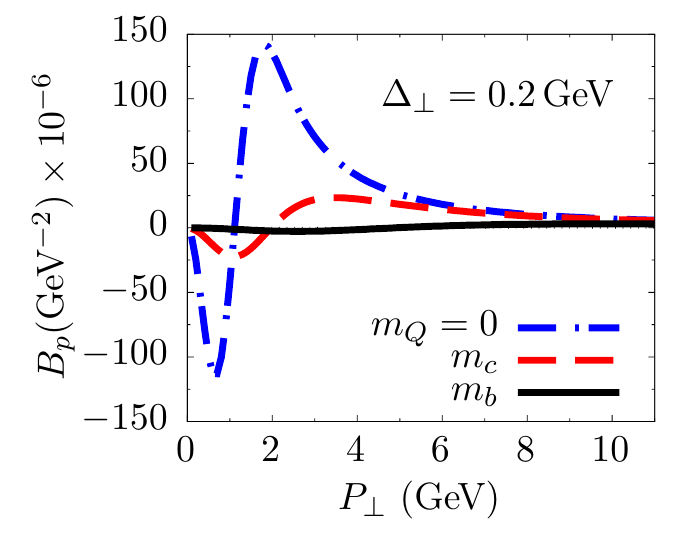} \hfill 
    \includegraphics[width=.48\textwidth]{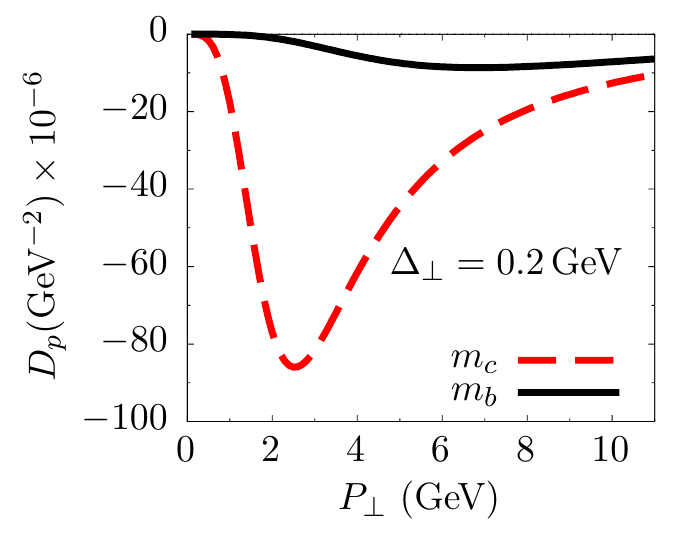}
    \caption{Left: structure function $B$. Right: structure function $D$. Both cases correspond to the proton target and are shown for massless and heavy ($c,b$) quarks as functions 
    of $P\p$ for a fixed $\Delta\p=0.2$ GeV.}
    \label{fig:proton:BD}
\end{figure}

Analogously, we show the structure functions for the proton target in Figs.~\ref{fig:proton:AC} and \ref{fig:proton:BD} for heavy ($c,b$) and massless quarks as functions of the relative quark-antiquark momentum $P\p$ for $\Delta\p=0.2$ GeV. These are also new results, but now regarding the proton structure. The proton has structure functions with smaller absolute values than the nucleus as our results are not normalized by the number of nucleons. 

The absolute value of structure functions $C$ and $A$ are again comparable, while in here $D$ is even larger than $B$. The behaviour w.r.t.\ the quark masses is analogous to the nucleus case.

In contrast to the nucleus, the MV model for the proton target presents a smaller difference in the $A$ and $C$ structure functions for different $\Delta\p$, i.e. just a small decrease in the absolute value and no sign flip within the kinematic range considered here. That is why in this case we chose not to show any figures with different $\Delta\p$, as such a behavior can be directly inferred from Fig.~\ref{fig:T-proton}. 

Also for larger $\Delta\p$, the elliptic structure functions $B$ and $D$ become even more relevant than in the nucleus case. Comparing Figs.~\ref{fig:T-nucleus} and \ref{fig:T-proton}, we see that they will increase with $\Delta\p$ in the proton case, while in the nucleus case they decrease. Of course, this depends on the $\Delta\p$ range we work with. If very large, the proton structure functions may show a similar behavior. Our choices of $\Delta\p$ are reasonable within the detector constraints available at the LHC.

\subsection{Exclusive quark-pair photoproduction cross sections}

Now we present our cross section results. We calculate the hadron cross section integrated in angle with exact kinematics. However, it is instructive to understand what happens in the limit $k_{1, 2 \perp} \rightarrow P\p$. In this limit, azimuthal integration produces terms proportional to $2A^2 + B^2 $ or $2C^2 + D^2$. We numerically investigated this approximation and found that it has negligible impact on the final result for our choice of small $\Delta\p$.
\begin{figure}[!h]
    \centering
    \includegraphics[width=.48\textwidth]{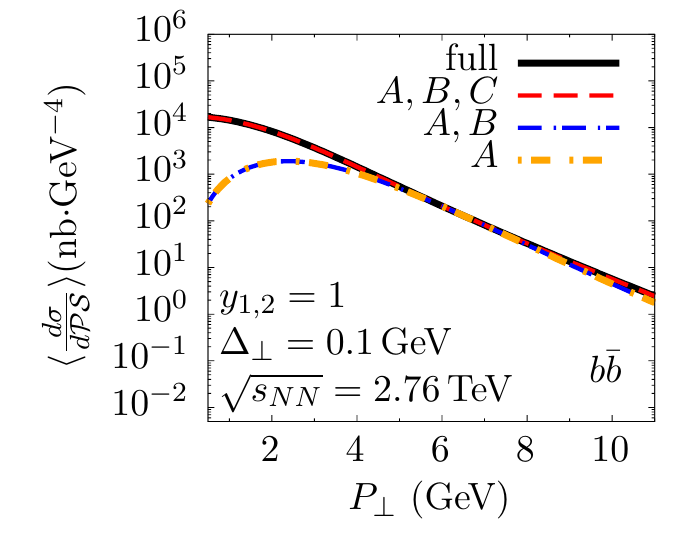} \hfill
    \includegraphics[width=.48\textwidth]{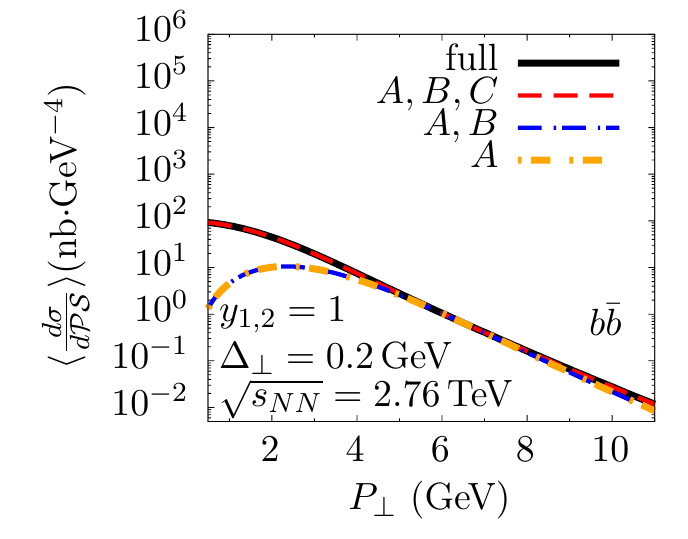}
    \caption{The fully differential $b\bar b$ photoproduction cross section in lead--lead UPCs as a function of the quark-antiquark transverse momentum difference $P\p$. The azimuthal angles are integrated. Here, {\it full} means all terms in the cross sections were considered (solid line), while $A,B,C$ means $D$ was set to zero (dashed line), and so on. In the left (right) plot $\Delta\p = 0.1$ GeV ($0.2$ GeV).}
    \label{fig:AvgSigma_bb}
\end{figure}

For all plots we chose the quarks rapidities $y_{1,2}$ equal to 1. In this way, we work in the forward region accessible by ATLAS and CMS (which also justifies our choices of c.o.m.\ energies). This choice is justified also by the fact that the MV model is fitted for low $x$. Therefore, the forward moving nucleus will provide the photons while the target (backward moving) will provide the small $x$ gluons. 
\begin{figure}[htb]
    \centering
    \includegraphics[width=.48\textwidth]{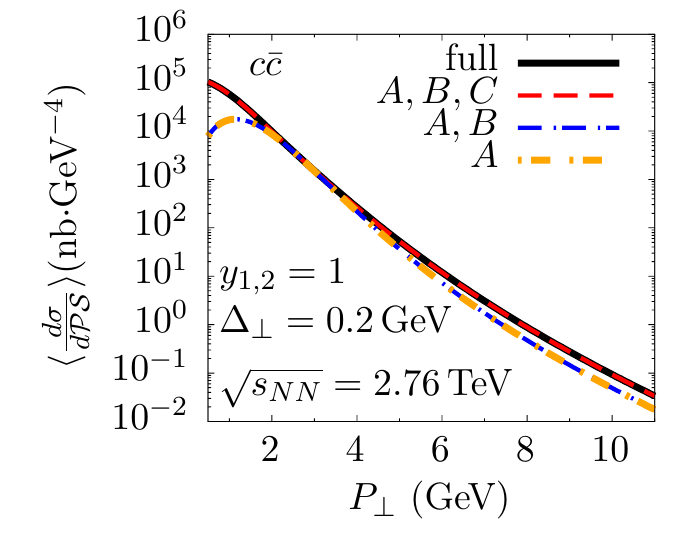}
    \caption{The same plot as in Fig.~\ref{fig:RatioDt_bb} but for $c\bar c$ production.}
    \label{fig:AvgSigma_cc}
\end{figure}

First, we present our cross section results for $b\bar b$ production in lead--lead UPCs, where $A = 208$ and $Z = 82$. The hadron cross section is depicted in Fig.~\ref{fig:AvgSigma_bb} as a function of $P\p$. Two plots are shown for $\Delta\p = 0.1$ and $0.2$ GeV, on left and right panels, respectively. The behaviour at large $P\p$ resembles an exponential decay and is expected because MV model does not take into account QCD evolution.
\begin{figure}[htb]
    \centering
    \includegraphics[width=.45\textwidth]{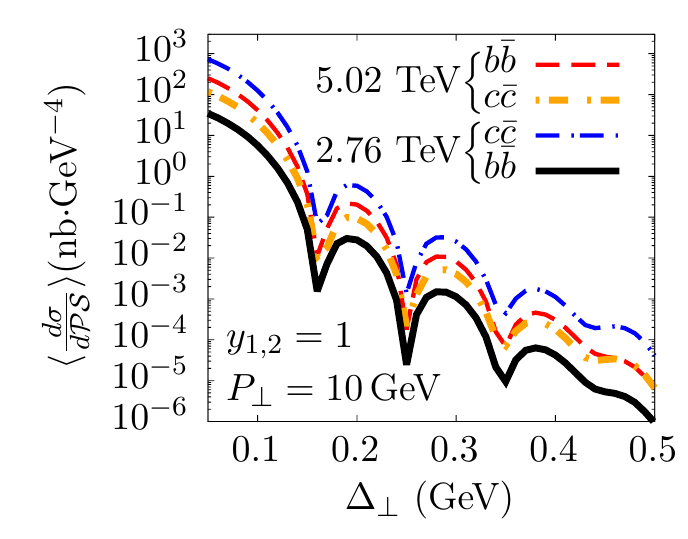}
    \caption{The fully differential cross section of $b\bar b$ 
    and $c\bar c$ pair photoproduction in lead--lead UPCs as a function of the final target transverse momentum $\Delta\p$. 
    The azimuthal angles are integrated. The dips are not 
    affected by a change in c.m. energy.}
    \label{fig:sgmAvg}
\end{figure}

When looking closer at the $\Delta\p$ variation, it is seen that there is no substantial difference in the shape of the cross section other than it is roughly divided by a factor of the order of 100. In the previous section, it was shown that in going from $\Delta\p = 0.1$ to $0.2$ GeV the sign of all the structure functions flips and their magnitudes reduces by a factor of the order of 10. As the cross section is dominated by $A^2$, the sign is irrelevant. The observation that there is a larger cross section for smaller $\Delta\p$ is just a result of the fact that the quark pair will more likely be created in a back-to-back configuration.

In Fig.~\ref{fig:AvgSigma_bb} we also show what happens if one sets to zero some of the structure functions. For instance, with $C=D=0$, the elliptic part $B$ plays an insignificant role compared to the $A$ contribution. The same can be said about the other elliptic structure function $D$. However, when $C$ is turned on, there is a small but significant difference at $P\p\approx 10$\,GeV and a relatively large contribution at small $P\p$. Therefore, the contribution proportional to $C^2$ can be directly measured with an appropriate choice of kinematical cuts.
\begin{figure}[htb]
    \centering
    \includegraphics[width=.48\textwidth]{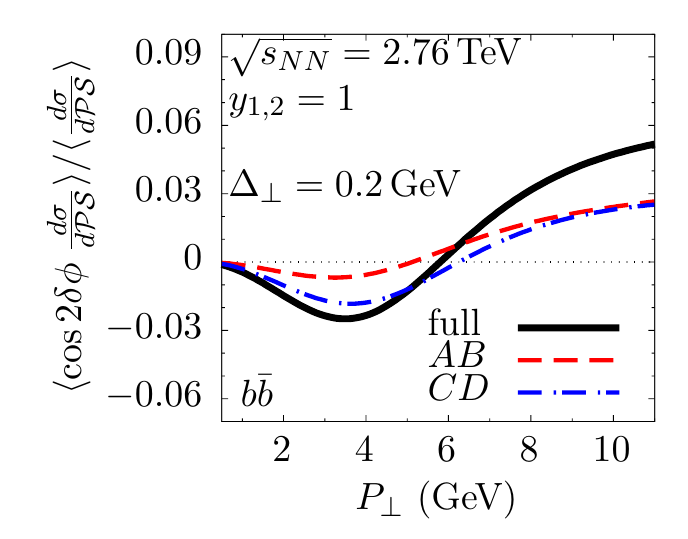} \hfill
    \includegraphics[width=.48\textwidth]{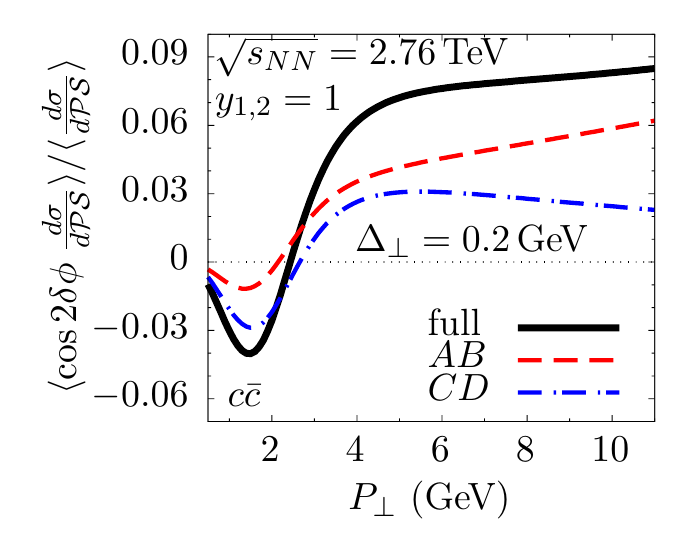}
    \caption{The ratio between the angular cosine-weighted average
    over the angle integrated cross section as a function 
    of $P\p$ with fixed $\Delta\p = 0.2$ GeV in the case of 
    lead target. Left: bottom 
    quark case; 
    right: charm quark case.}
    \label{fig:RatioDt_bb}
\end{figure}

In Fig.~\ref{fig:AvgSigma_cc} charm production in Pb+Pb collisions for $\Delta\p=0.2$ GeV is depicted. Overall, it has the same features as for bottom production, but it is a larger cross section, especially at small $P\p$ when it becomes a factor $10^3$ larger. Another point is that at $P\p \approx 10$\,GeV the relative size of the $C$ structure function contribution is larger when compared to the bottom quark case.

So it appears that charm quark pair production would be a better observable than bottom quark-pair for our purposes. It has a much higher cross section and it discerns the $C$ structure function better. However, there is an important issue: quark-meson fragmentation. The detectors will be able to measure transverse momentum of $D$ and $B$ mesons only, i.e., they do not access quark level variables. In spite of that, for a large heavy quark mass, there is a well known effect: the produced meson will have most of the momentum of the heavy quark. Therefore, the bottom quark observables have an advantage that fragmentation does not washes away too much the momentum distributions~\cite{Proceedings:khoze}. For the charm quark case, the $c\to D$ fragmentation may be relevant for a comparison with future measurements. But since the corresponding calculations become very difficult to perform numerically, and their theoretical interpretation becomes less transparent, we leave this point for future studies.
\begin{figure}[htb]
    \centering
    \includegraphics[width=.48\textwidth]{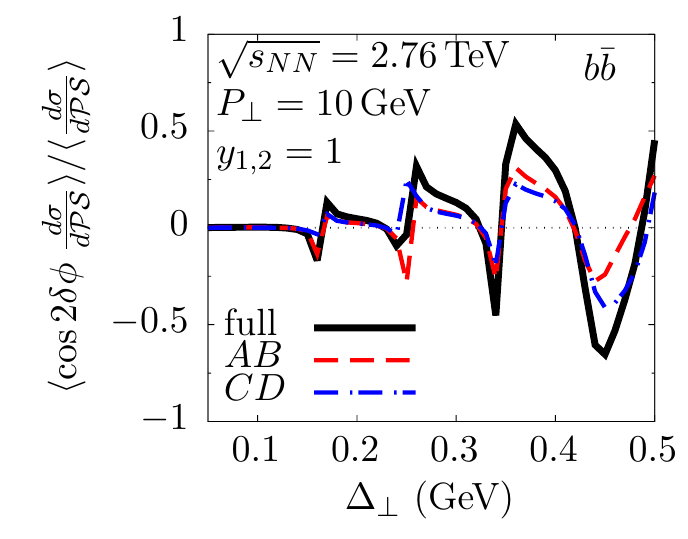} \hfill 
    \includegraphics[width=.48\textwidth]{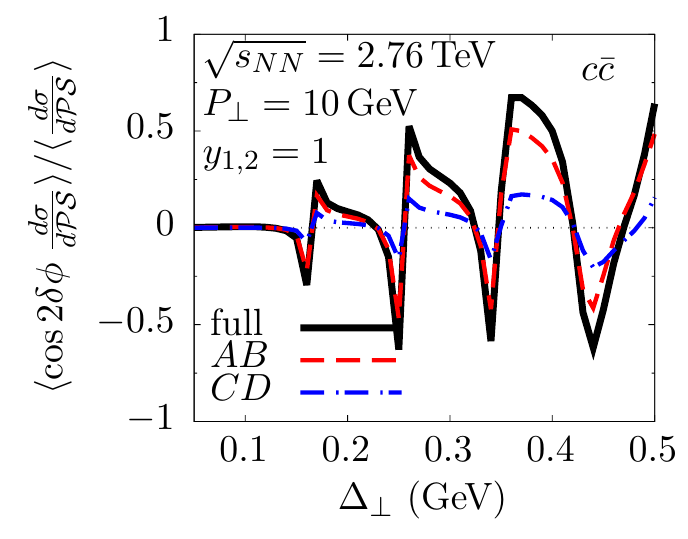}
    \caption{The ratio of the angular cosine-weighted average
    to the angle integrated cross section as a function 
    of $\Delta\p$ with fixed $P\p = 10$ GeV in the case of 
    lead target. Left: bottom quark 
    case; right: charm quark case.}
    \label{fig:RatioPt_bb}
\end{figure}

The next Fig.~\ref{fig:sgmAvg} shows the same $b\bar b$ and $c\bar c$ production cross sections but as functions of the target final transverse momentum $\Delta\p$. The variable $P\p$ is fixed at 10\,GeV. We remark that the dips (minima) are not affected by changing the c.m. energy. They are a direct result of the oscillations in MV model, whose scale is related to the saturation scale. It is still important to state that they also do not depend on whether bottom or charm quarks are being produced, and thus exhibit important probes for the proton or nucleus structure.

At this point we turn our attention to a different observable. As seen above, the angular-integrated cross sections discussed above are not very convenient for getting any physics information about the elliptic part. Therefore, instead we would like use the cosine-weighted angle average determined as follows:
\begin{equation} \label{eq:cosine} 
\left\langle \frac{d\sigma ^{p A} }{d\mathcal{PS}}  \cos 2(\phi_P - \phi_\Delta) \right\rangle  = \int_0^{2 \pi} d\phi_{P\p}  \int_0^{2 \pi} d\phi_{\Delta\p}  \frac{d\sigma ^{p A} }{d y_1 d y_2 \, d^2 \vec P\p \, d^2 \vec \Delta\p}  \cos 2(\phi_P - \phi_\Delta) 
\end{equation}
Roughly speaking, the more positive this observable is, the more $P\p$ and $\Delta\p$ are parallel (or antiparallel); the negative case correlates with perpendicular vectors.
\begin{figure}[htb]
    \centering
    \includegraphics[width=.48\textwidth]{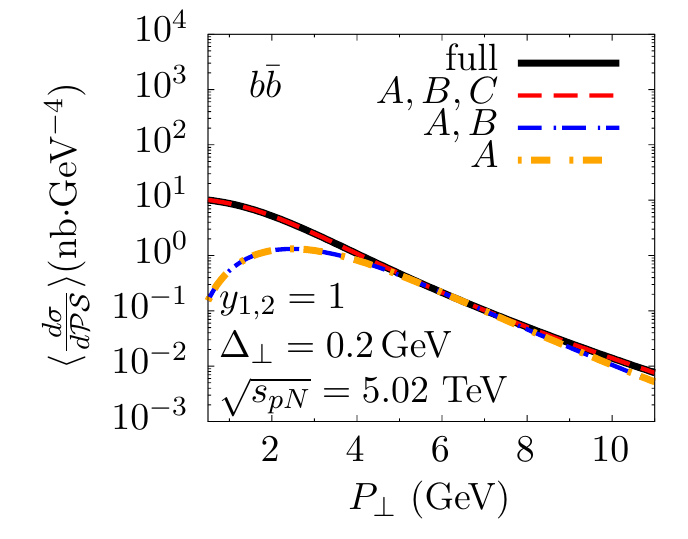} \hfill \includegraphics[width=.48\textwidth]{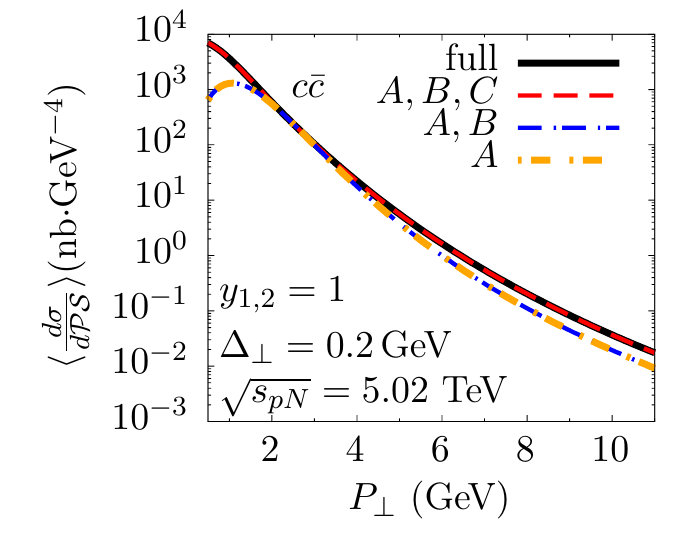}
    \caption{
    The fully differential cross section of $b\bar b$ (left) and $c\bar c$ (right) pair photoproduction in lead--proton UPCs as a function of the quark-antiquark relative transverse momentum $P\p$, with quark and antiquark rapidities $y_{1,2} = 1$.}
    \label{fig:AvgSigma_pA}
\end{figure}

An azimuthal angle distribution is relatively easy to measure, besides not being influenced by fragmentation. Also, Eq.~\ref{eq:cosine} is a ratio of the cross sections implying largely reduced theoretical and experimental uncertainties. For instance, the luminosity as well as the overall normalization are canceled out in such a ratio. We conclude that this is a very reliable observable.
\begin{figure}[htb]
    \centering
         \includegraphics[width=.48\textwidth]{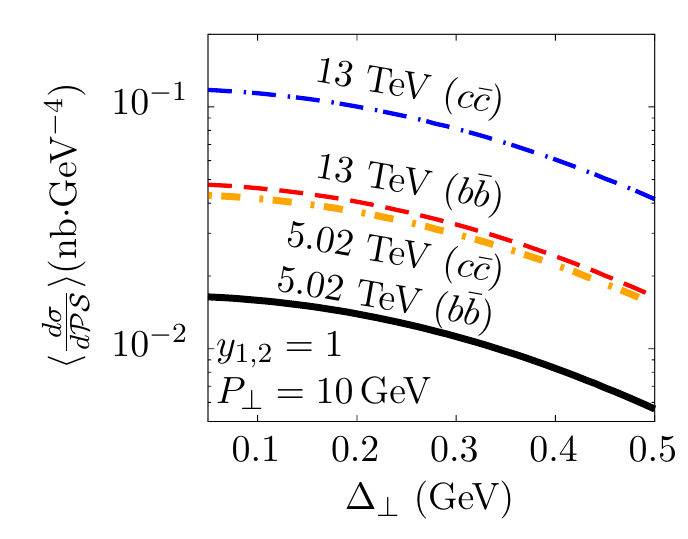}
    \caption{
    The fully differential cross section of $b\bar b$ 
    and $c\bar c$ pair photoproduction in lead--proton UPCs as a function of the final proton transverse momentum $\Delta\p$. 
    The azimuthal angles are integrated.}
    \label{fig:avg_pA}
\end{figure}

Again, we use exact kinematics, but if we set the quark momenta to be equal $P\p$, the integration over the differential cross section multiplied by $\cos 2( \phi_P - \phi_\Delta)$ lets only terms $A\cdot B$ or $C\cdot D$ survive. As such, the study of this observable is relevant to determine the size of the elliptic contribution, and to determine which kinematic region is more interesting to obtain phenomenological information on the Wigner distribution.
\begin{figure}[htb]
    \centering
    \includegraphics[width=.48\textwidth]{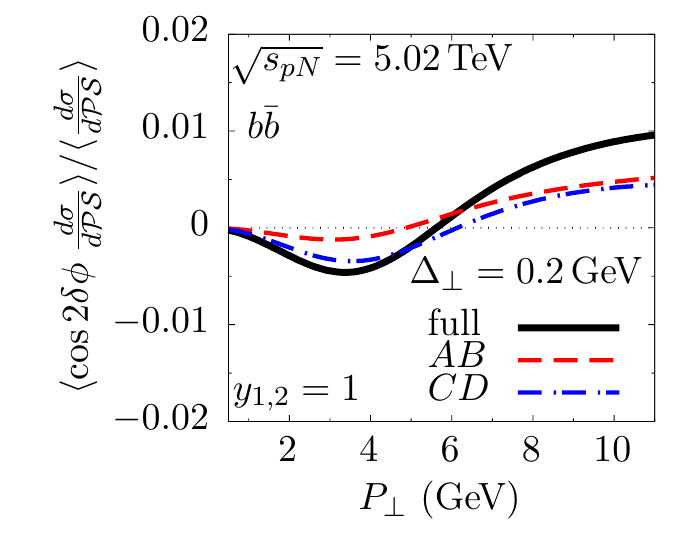}\hfill
     \includegraphics[width=.48\textwidth]{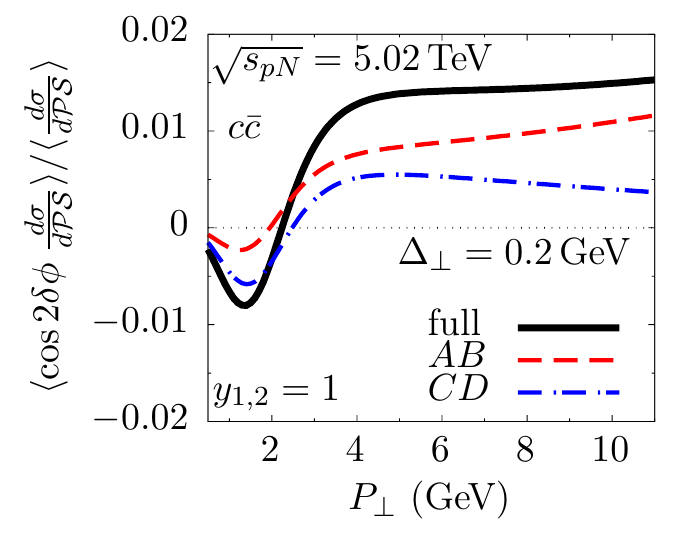}
    \caption{
    The ratio between the angular cosine-weighted average
    over the angle integrated cross section as a function 
    of $P\p$ with fixed $\Delta\p = 0.2$ GeV in the case of 
    proton target. Left: bottom quark case; right: 
    charm quark case.}
    \label{fig:RatioDt_pA}
\end{figure}

The ratio of the cosine-weighted angle integrated over the angular-integrated cross section as a function of $P\p$ is shown in Fig.~\ref{fig:RatioDt_bb}. We show both the $b \bar b$ and $c \bar c$ production cross sections, with fixed $\Delta\p = 0.2$\,GeV. We see that with appropriate choices of $P\p$ the information can be extracted about both $B$ and $D$ structure functions.

It is expected from Figs.~\ref{fig:A} and \ref{fig:B} that the contribution from the elliptic term rises as $\Delta\p$ increases. This can be seen better by fixing $P\p$ and varying $\Delta\p$, as in Fig.~\ref{fig:RatioPt_bb}. Besides, we see an oscillation as expected.
\begin{figure}[htb]
    \centering
         \includegraphics[width=.48\textwidth]{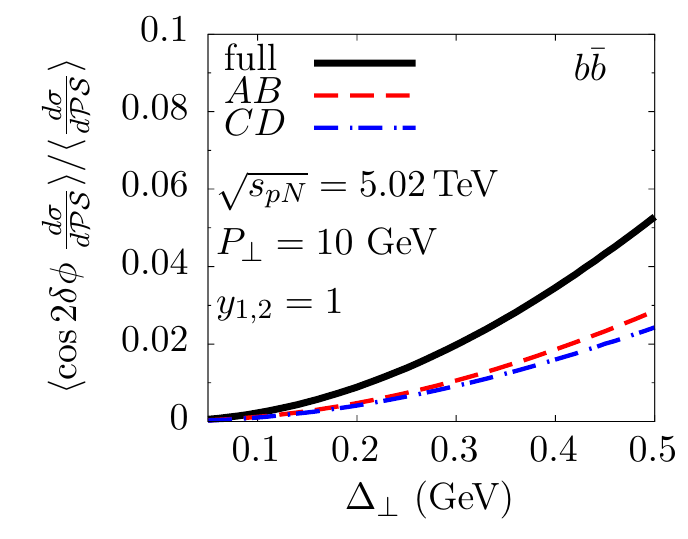} \hfill
         \includegraphics[width=.48\textwidth]{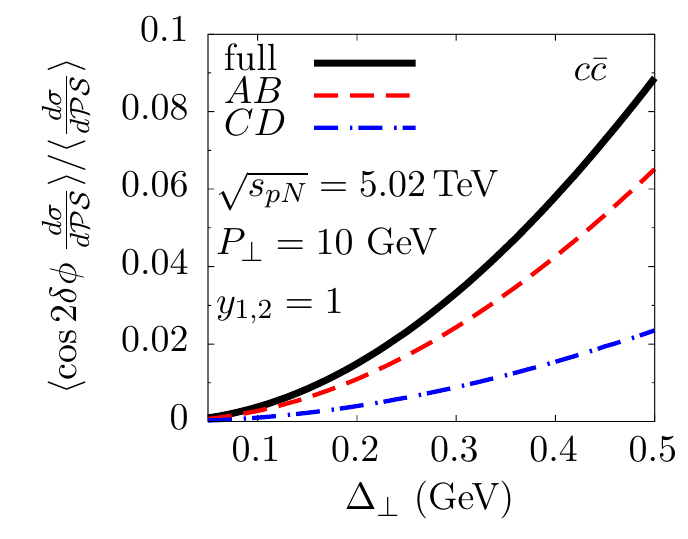}
    \caption{Ratio of angle integrated cosine-weighted average over angle integrated cross section with the proton as a target as a function of $\Delta\p$ with fixed $P\p = 7.5$ GeV. Open $b \bar b$  ($c \bar c$) production in the left (right).}
    \label{fig:RatioPt_pA}
\end{figure}

What would change if the target were a proton? In Figs.~\ref{fig:AvgSigma_pA}--\ref{fig:RatioPt_pA}, we show this case as well. Of course the cross sections are smaller since in the heavy ion case we did not divide by $A$. The dependence on $P\p$ is pretty much the same as in the nuclear case. However, the dependence on $\Delta\p$ does not show oscillations for the ranges studied. For instance, the  cosine-weighted average increases steadily with $\Delta\p$ as is seen in Fig.~\ref{fig:RatioPt_pA}.

\section{Conclusions}
\label{Sect:Concl}

In this work we have analyzed exclusive heavy quark photoproduction in the forward region in pA and AA UPCs with quasi-real WW photons as a mean to constrain the gluon Wigner distribution. Our new results can summarized in three main points. 

First, we derived the analytic expressions at leading order for the calculation of the observable. In doing so, we introduced two new structure functions $C(P\p,\Delta\p)$ and $D(P\p,\Delta\p)$ due to the fact that the quarks have mass, besides adding mass corrections to the already known $A(P\p,\Delta\p)$ and $B(P\p,\Delta\p)$. By using the generalized MV model for lead and proton targets, we studied the non-trivial features of these functions, especially the ones related to the elliptic component of the Wigner distribution, $B$ and $D$. These are of definite importance to understand the angular correlations between the transverse momenta $k\p$ and $\Delta\p$ in the GTMD, and can be related to elliptic flow in hadron and/or nuclei collisions.

Secondly, we numerically calculated the bottom and charm pair production cross sections in a fully differential form using the above structure functions, both for proton and nucleus target.  We provided results in the forward region, which is accessible by ATLAS and CMS. Regarding the azimuthal-angle integrated cross section, we showed that the new $C$ structure function has a big impact in the low-$P\p$ region ($P\p\lesssim 4$\,GeV), while $A$ dominates the differential cross section in the range $4\lesssim P\p\lesssim7$\,GeV. The $C$ function starts to be relevant again for $P\p \gtrsim 7$\,GeV. We also compared heavy and light quark results, and as a side effect we calculated pA cross section for light quarks using the  Iancu-MV model, not done before to the best of our knowledge. 

Finally, in a more phenomenological minded investigation, we defined the cosine-weighted angular average of the differential cross section in order to access the elliptic part of the hadron structure. This special average can help to constrain the elliptic distribution, since it is directly connected to the products $A\cdot B$ and $C \cdot D$. As neither of the products dominates in the considered kinematic regions, they can be be probed simultaneously by a measurement.

Having these new results, we want to state again why they are important. From the phenomenological point of view, the study of heavy-quark di-jets is relevant in comparison to its light quark equivalent, since it is less affected by fragmentation effects and has a cleaner QCD background. Also, it has much smaller theoretical uncertainties w.r.t. higher order corrections, as light quark jets suffer from potentially huge corrections.

From the experimental point of view, measurements can be done at lower transverse momenta then for jets emerging from the light quarks. The cross sections calculated here can have transverse momentum of the order of the saturation scale, where most of non-trivial features of the Wigner distribution are predicted. Such low $p\p$ has never been achieved before and can be technically reached by tagging on open heavy flavoured mesons. We hope that the observables of this paper will be taken into account by the forward physics experimental groups when planning for new measurements and analysis, since they are highly relevant and measurable, for instance, at ATLAS, CMS and LHCb. 

\section*{Acknowledgments}

Useful discussions with Y.~Hatta, M. Tasevsky, O. Teryaev at early stages of this work are gratefully acknowledged. R.P.~is supported in part by the Swedish Research Council grants, contract numbers 621-2013-4287 and 2016-05996, by CONICYT grants PIA ACT1406 and MEC80170112, as well as by the European Research Council (ERC) under the European Union's Horizon 2020 research and innovation programme (grant agreement No 668679). This work was supported by Fapesc, INCT-FNA (464898/2014-5), and CNPq (Brazil) for EGdO and MRP. This study was financed in part by the Coordena\c{c}\~ao de Aperfei\c{c}oamento de Pessoal de N\'ivel Superior -- Brasil (CAPES) -- Finance Code 001. The work has been performed in the framework of COST Action CA15213 ``Theory of hot matter and relativistic heavy-ion collisions'' (THOR). This work was also supported in part by the Ministry of Education, Youth and Sports of the Czech Republic, project LT17018.

\bibliography{bib}

\end{document}